\documentclass{aa}
\usepackage{epsfig}

\def\rerg{\rm erg}
\def\rs{\rm s}
\def\rs1{\rm s^{-1}}

\def\rcm{\rm cm}
\def\rcm2{\rm cm^{-2}}
\def\deg{\rm ^{\circ}}
\def\flux{\rerg\ \rcm2\ \rs1}

\def\flue{\rerg\ \rcm2}
\def\rkeV{\rm keV}

\def\cqr{\chi^2_\nu}
\def\c2nor{\chi^2}
\def\ga{\mathrel{\hbox{\rlap{\hbox{\lower4pt\hbox{$\sim$}}}\hbox{$>$}}}}
\def\la{\mathrel{\hbox{\rlap{\hbox{\lower4pt\hbox{$\sim$}}}\hbox{$<$}}}}
\def\sax{{\it BeppoSAX}}
\def\chandra{{\it Chandra}}
\def\hete2{{\it HETE-2}}

\begin{document}


\title{Prompt and afterglow X--ray emission from the X--Ray Flash of 2002 April 27}

\author{L.~Amati\inst{1}
\and F.~Frontera\inst{1,2}
\and J.J.M.~in 't Zand\inst{3,4}
\and M.~Capalbi\inst{5}
\and R.~Landi\inst{1}
\and P.~Soffitta\inst{6}
\and L.~Vetere\inst{6}
\and L.A.~Antonelli\inst{7}
\and E.~Costa\inst{6}
\and S.~Del Sordo\inst{8}
\and M.~Feroci\inst{6}
\and C.~Guidorzi\inst{2}
\and J.~Heise\inst{3,4}
\and N. Masetti\inst{1}
\and E.~Montanari\inst{2}
\and L.~Nicastro\inst{8}
\and E.~Palazzi\inst{1} 
\and L.~Piro\inst{6} 
}

\offprints{L. Amati:amati@bo.iasf.cnr.it}

\institute{Istituto di Astrofisica Spaziale e Fisica cosmica - Sezione di Bologna,
CNR, Via Gobetti 101, I-40129 Bologna, Italy
\and
Dipartimento di Fisica, Universit\`a di Ferrara, Via Paradiso
 12, I-44100 Ferrara, Italy
\and
SRON National Institute for Space Research,
 Sorbonnelaan 2, 3584 CA Utrecht, The Netherlands
\and
Astronomical Institute, Utrecht University, P.O. Box 80000, 
3508 TA Utrecht, The Netherlands
\and
ASI Science Data Center c/o ESRIN, Via G. Galilei,
  I-00044 Frascati (RM), Italy
\and
Istituto di Astrofisica Spaziale e Fisica cosmica, CNR, Via Fosso del Cavaliere,
  I-00133 Roma, Italy
\and
Osservatorio Astronomico di Roma, Via Frascati 33,
  I-00040 Monteporzio Catone (RM), Italy
\and
Istituto di Astrofisica Spaziale e Fisica cosmica - Sezione di Palermo, CNR,
Via La Malfa 153, I-90146 Palermo, Italy
}

\date{Received; Accepted }

\markboth{Prompt and afterglow X--ray emission
from XRF~020427. 
}{}

\abstract{ 
 We report on the X--ray observations of the X--ray flash (XRF) which
occurred on 2002 April 27, three days before \sax\ was switched off. The event
was detected with the \sax\ Wide Field Cameras but not with the Gamma ray Burst 
Monitor. A follow-up observation with the \sax\ Narrow Field Instruments was 
soon performed
and a candidate afterglow source was discovered. We present the results obtained.
We also include the results obtained from the observations of the
XRF field with the \chandra\ X--ray satellite. The spectral analysis of the prompt 
emission shows that the peak energy of the $EF(E)$ spectrum is lower than 
5.5~keV, with negligible spectral evolution.
The X--ray afterglow spectrum  is consistent with a power law model with
photon index of $\sim 2$, while the 2--10 keV flux fades as a power law
with a decay index $-1.33$. Both these indices are
typical of GRBs. 
A very marginal excess at $\sim$4.5--5 keV is found in the afterglow 
spectrum measured by \sax\ .
As for many GRBs, the extrapolation of the 2--10 keV 
fading law back to the time of the prompt emission is consistent 
with the X--ray flux measured during the second part of the event.
We estimate a possible range of values of the redshift and 
discuss our results in the light of current models of XRFs. 
}
\authorrunning{Amati et al.}
\titlerunning{Prompt and afterglow X--ray emission
from XRF~020427}{}

\maketitle


\section{Introduction}
\label{intro}

Among the Fast X--ray Transients (FXTs) detected by the \sax\ Wide Field Cameras 
(WFC, 2--28~keV, \cite{Jager97}) on--board the \sax\ satellite (\cite{Boella97a}) 
in six years of operation, there have been more than 20 events 
(\cite{Heise03}) whose temporal and 
spectral properties resemble those of the X--ray counterparts of Gamma--Ray 
Bursts (GRBs), but were not detected by the  Gamma--Ray Burst Monitor 
(GRBM, 40--700~keV, \cite{Frontera97}). They are called
X--Ray Flashes (XRFs). The classification as XRF is commonly extended to 
GRB--like events detected by HETE--2 showing no signal above $\sim$30 keV in the 
FREGATE instrument (e.g. \cite{Barraud03}). Distinctive features of XRFs with 
respect to the other FXTs are their shorter duration (from a few tens to a few 
hundreds of seconds), their light curves, their 
non--thermal and quickly evolving spectra, and their isotropic distribution 
in the sky (e.g. \cite{Heise03}). The origin of these events is a matter of debate:
either they have a nature completely different from GRBs or they are very 
X--ray rich GRBs, too soft to be detected 
by the utilized gamma--ray instruments. 
The second interpretation seems to be
more consistent with the data. In fact, the extrapolation to higher photon 
energies of the WFC spectra of these 
events indicates that for all but two XRFs the expected signal in the 40--700~keV 
energy band is below the GRBM sensitivity threshold 
(\cite{Heise03}). Also, an inspection of the CGRO/BATSE light curves in 
the 25--50~keV and 50--100~keV energy 
bands resulted in a positive detection ($>$5 $\sigma$) for 9 out of 10 XRFs 
observed with the WFCs (\cite{Kippen03}). The one remaining case may very
well be a thermonuclear flash on a Galactic neutron star (\cite{Cornellisse02}).
The spectral analysis of a sample 
of 35 XRFs detected by \hete2\ (\cite{Barraud03}) further supports the 
interpretation of XRFs as very soft GRBs,
with peak energies of the $EF(E)$ spectra which can be as low as 
a few keV, and with peak fluxes or fluences much higher in X--rays than
in gamma--rays. 
Various explanations of the X--ray richness of these events have been 
proposed,
for example fireballs with low Lorentz factors due to a high baryon loading 
(e.g. \cite{Dermer99,Huang02}), very high redshift GRBs ($z>$5) 
(e.g. \cite{Heise03}), collimated GRBs seen at large off axis angles  
(e.g. \cite{Yamazaki02}).    

The increasing number of XRF observations with \hete2 is providing important 
information
for the clarification of the phenomenon. The assumption of high redshift GRBs
does not seem to be confirmed by the data. 
The detection of X--ray afterglow emission from XRF~011030 with \chandra\
(\cite{Harrison01}) was followed by the discovery of a fading radio 
counterpart (\cite{Taylor01}) and of a blue host galaxy possibly located 
at z$<$3.5 (\cite{Bloom03}). 
The follow--up observation of XRF~030723 (\cite{Prigozhin03}),
led to the discovery of optical (\cite{Fox03}) and X--ray (\cite{Butler03}) 
afterglow counterparts. The spectra of the optical transient (OT) of 
this event suggest a 
redshift lower than 2.3 (\cite{Fynbo04}). The claim of an optical and radio 
counterpart of XRF~020903 (\cite{Berger02,Soderberg02,Soderberg03}) and its 
possible 
association with a SN event at  $z = 0.25$ is debated: based on
inspections of the available optical and radio data of 
the field, Gal--Yam et al. (2002) \nocite{GalYam02} suggest that the variable 
object could be a radio--loud AGN. 

In this paper, we present and discuss the prompt and delayed emission properties
of XRF~020427 observed with the \sax\ /WFC and GRBM instruments and followed-up
with both \sax\  and \chandra\ satellites.

\begin{center}
\begin{table*}[ht!]
\caption{Log of the X--ray observations of XRF~020427}
\begin{tabular}{llllll}
\hline
\hline Instrument & Band &  Seconds from  & Exposure & Source counts & Average 2--10 $\rkeV$ flux$^{(a)}$ \\
    &  ({\rm keV}) & XRF onset  & (s)  &  & ($\flux$ )   \\
\hline
\sax\ /WFC & 2--28  & 0--60 & 60 & 4774$\pm$329 & (9.7$\pm$0.7)$\times$10$^{-9}$ \\ 
\sax\ /MECS & 1.6--10 & 40320--55080 & 6834 & 41$\pm$9 & (4.2$\pm$0.9)$\times$10$^{-13}$ \\
\sax\ /MECS & 1.6--10 & 197640--218520 & 8044 & $<$24$^{(b)}$ & $<$2.1$\times$10$^{-13}$$^{(b)}$ \\
\hline
\chandra\ /ACIS--S$^{(c)}$ & 0.3--7  & 770503--786408 & 13741 & 55$\pm$8 & (1.1$\pm$0.2)$\times$10$^{-14}$ \\
\chandra\ /ACIS--S$^{(c)}$ & 0.3--7  & 1453875--1470117 & 14568 & 23$\pm$6 & (4.4$\pm$1.2)$\times$10$^{-15}$ \\
\hline
\hline
\end{tabular}
\begin{list}{}{}
\item[$^{(a)}$] Values computed by assuming the best fit spectral models
(see Table~\ref{t:tab2} and text)
\item[$^{(b)}$] 3$\sigma$ upper limits.
\item[$^{(c)}$] The values here reported refer to CXOU J220928.2--651932 and
were derived from the \chandra\ public data archive (see also Sect.~3). 
\end{list}
\label{t:tab1}
\end{table*}
\end{center}
\section{Observations}
\label{s:obs}

\begin{figure}
\psfig{figure=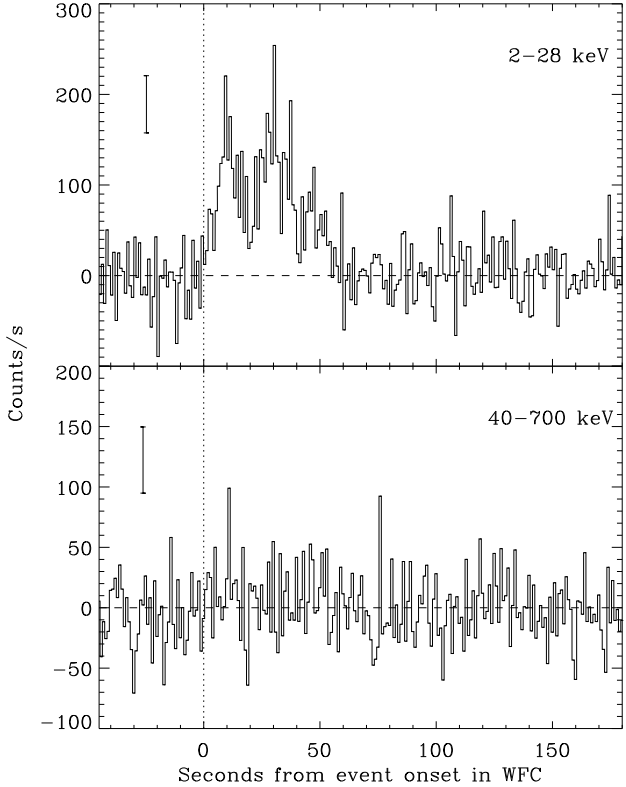,width=9cm}
\vspace{0.5cm}
\caption{Background subtracted light curve of XRF~020427 in the 
2--28 $\rkeV$ ({\it top}) and 40--700 $\rkeV$ ({\it bottom}) energy bands. 
The integration time of each bin is 1 s.}
\label{f:prompt_lc}
\end{figure}

XRF~020427 was detected by \sax\ /WFC unit 2 on 2002 April 27, at 
03:48:40 UT with no signal from the GRBM (Fig.~\ref{f:prompt_lc}). 
Simultaneously, a sudden ionospheric disturbance
lasting $\sim$60 s with an amplitude (and therefore ionizing flux) 
comparable to that measured on 1998 August 27 from the giant flare from
SGR1900+14
(\cite{Fishman02}). It is not clear whether the disturbance on 2002 April 27
is associated with the XRF, but if it is the XRF it should also have had a very 
intense flux below 2 keV. Unfortunately, there are no soft X--ray measurements
available.
XRF~020427 was localized with an uncertainty of 3' (at 99\% confidence).
It was classified as an X--ray 
Flash rather than a Galactic X--ray burst because it failed to exhibit 
spectral softening 
during the decay, is relatively far from the Galactic plane 
($-$44.2$^\circ$), and is not 
coincident with a known Galactic X--ray source (\cite{Zand02}). Two Target Of 
Opportunity (TOO) observations of the WFC error circle were performed with the 
\sax\ / Medium Energy Concentrator Spectrometer (MECS, 1.6--10~keV, 
\cite{Boella97b}). None of the other Narrow Field Instruments were switched on.
The TOOs lasted from 11.2 to 15.3~hrs and 
from 54.9 to 60.7~hrs after the XRF. A previously unknown X--ray source
was detected, 
designated 1SAX~J2209.3--6519, at 
$\alpha_{2000.0}
= 22^{\rm h}09^{\rm m}23^{\rm s}, \delta_{2000.0} = -65^{\rm
o}19^\prime34\arcsec$,
1.1' away from the centroid of the WFC error circle (\cite{Amati02}). 
Given that the flux of the source decreased by a factor of $\sim$2 from 
the first to the second TOO, it was identified as the likely X--ray afterglow 
of XRF~020427. 

\chandra\ observed XRF~020427 on 2002 May 6 and 14 with ACIS--S (0.3--10 keV,
\cite{Garmire03}) at the focal plane and with no grating. 
Three previously unknown sources were detected within the MECS--determined error circle
of radius 1$^\prime$ of which one (CXOU J220928.2--651932) faded 
substantially from the first to the second observation (\cite{Fox02}). 
One previously unknown radio source lying inside the MECS error circle was 
detected at 8.7 GHz with the Australian radio telescope ATCA  (\cite{Wieringa02}),
but none of the three \chandra\ sources had a position coincident
with that of this radio source. Optical observations with
HST on 2002 June 10 and with the
VLT one day later showed that the
\chandra\ source lies on the edge of a 1.5$''$ large galaxy with V magnitude 
$\sim 24.7$ belonging to a group 
of three blue galaxies (\cite{Castro--Tirado02,Fruchter02}). Based on a combined 
analysis of the HST and \chandra\ images, Bloom et al. (2003) infer an
upper limit for the redshift of 3.5. This limit was fine--tuned 
to 2.3 with Gemini South observations (\cite{Dokkum03}).
A summary of all X--ray observations of XRF~020427 is reported
in Table~\ref{t:tab1}.\\
We performed the \sax\ data analysis by using 
standard reduction techniques and software (see \cite{Jager97} for the WFC,
\cite{Amati99} for the GRBM and 
\cite{Boella97b} for the MECS). 
For the spectral analysis of the MECS data we used an image extraction radius 
of 3$'$ and the standard background spectrum.
The public \chandra\ data were reduced by using standard CIAO 2.2.1 procedures. 
The spectral analysis was performed by 
using XSPEC version 11.2 (\cite{Arnaud96}). 
The quoted errors and limits are given at a 90\% confidence level, except where 
otherwise noted.

%
%

%
%
\begin{figure}
\psfig{figure=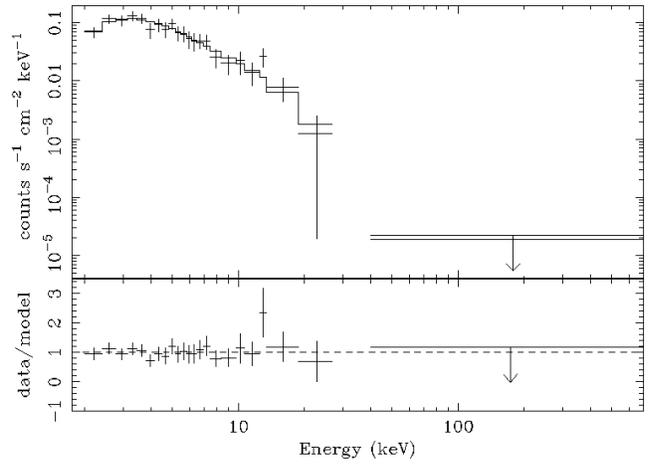,width=9.5cm}
\caption{Prompt emission spectrum of XRF~020427 fitted with 
a power law absorbed by the average Galactic column density along the
line of sight.}
\label{f:prompt_spec}
\end{figure}

\section{Results}
\label{s:res}

\subsection{Prompt emission}
\label{s:prompt}

The light curve of XRF~020427 in the 2--28~keV energy band
is shown in Fig.~\ref{f:prompt_lc} together with the simultaneous 
GRBM ratemeters in the 40--700~keV energy band. The X--ray event exhibits 
two pulses and a smooth decay with a $\sim$60 s duration, which is typical of
long GRBs. No signal is detected in gamma--rays. However
an excess in the counts at a confidence level of $2.7 \sigma$ appears 
when we compare
the total 40--700 keV counts measured during the X--ray event with the
background level.

The 2--28 $\rkeV$ average spectrum of XRF~020427 is shown in 
Fig.~\ref{f:prompt_spec}. It  
is well described ($\c2nor$/dof = 9.6/26) by a simple power law 
absorbed by the average Galactic hydrogen column density along the XRF direction 
($N_{\rm H}^G = 2.9 \times 10^{20}$~cm$^{-2}$), derived from radio maps 
(\cite{Dickey90}), and with
photon index $\Gamma$ = $2.09_{-0.21}^{+0.23}$. The extrapolation of this model
to higher energies is consistent with the 3$\sigma$ upper limit to the
40--700~keV event intensity for $\Gamma$$\ge$2.05. By fitting the WFC data with 
the Band function (\cite{Band93}) with $\alpha$ frozen to $-1$ and the other
spectral parameters ($E_{\rm 0}$, $\beta$ and normalization) free to vary, 
we get $\beta = -2.10_{-0.26}^{+0.22}$ and $E_{\rm 0} = 2.8_{-2.8}^{+2.7}$, 
with an upper limit 
of 5.5~keV to the peak energy E$_{\rm p}$ of the $EF(E)$ spectrum. The same 
result on $E_{\rm p}$ is obtained by fixing $\alpha$ at lower values and 
requiring the 
GRBM 3$\sigma$ upper limit to be satisfied. No 
spectral evolution during the event is seen in 
time resolved spectra.

%
%
\begin{figure*}
\psfig{figure=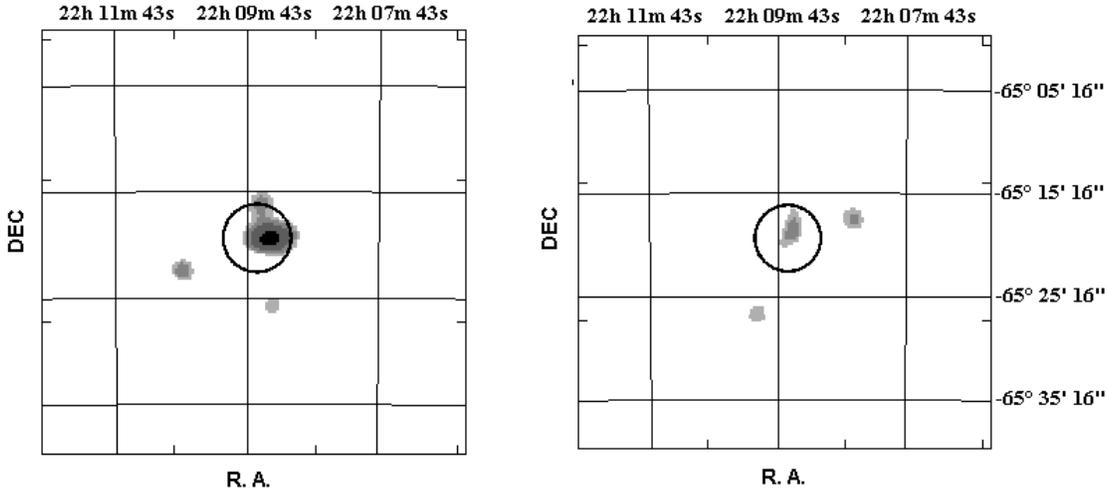,width=7.0cm,angle=-90}
\caption{\sax\ /MECS images of the field of XRF~020427 accumulated during
TOO~1 ({\it left panel}) and TOO~2 ({\it right panel}) 
in the  1.4--10 keV energy band. The error circle determined with the WFC 
is also shown in each image. This contains a source fading 
by a factor $\ga 2$
from the 1st to the 
2nd TOO observation.}
\label{f:mecs_image}
\end{figure*}

From the spectral results, we derive 1~s peak fluxes,
$F_{\rm 2-10 ~keV} = (1.9 \pm 0.3) \times 10^{-8}$$~\flux$ and 
$F_{\rm 2-28 ~keV} = (3.0 \pm 0.4) \times 10 ^{-8}$~$\flux$, and fluences, 
$S_{\rm 2-10 ~keV} = (3.7 \pm 0.3) \times 10 ^{-7}$~$\flue$
and $S_{\rm 2-28 ~keV} = (5.8 \pm 0.4) \times 10 ^{-7}$~$\flue$, which
lie in the range found for GRBs detected with the WFCs and GRBM 
(\cite{Frontera00b, Amati02}).
Assuming the spectral model which best fits the WFC data,
we also derive from the GRBM data
3$\sigma$ upper limits 
to the 1~s peak fluxes and fluences in gamma--rays: $F_{\rm 40-700 ~keV} 
< 6.6 \times10^{-8}$~$\flux$, $F_{\rm 50-300 ~keV} < 4.2 \times10^{-8}$~$\flux$, 
$S_{\rm 40-700 ~keV} <4.8 \times 10^{-7}$~$\flue$, 
$S_{\rm 50-300 ~keV} <3.1 \times 10^{-7}$~$\flue$. 
From the X--ray values and the GRBM upper limits we derive
the following $3 \sigma$ lower limits to the commonly used indicators of the
event X--ray richness (e.g. \cite{Feroci01,Kippen03}): 
$F_{\rm 2-10 ~keV}/F_{\rm 40-700 ~keV}$~$>$~$0.29$, \\
$F_{\rm 2-28 ~keV}/F_{\rm 40-700 ~keV}$~$>$~$0.45$, \\
$F_{\rm 2-10 ~keV}/F_{\rm 50-300 ~keV} > 0.45$, \\
$S_{\rm 2-10 ~keV}/S_{\rm 40-700 ~keV}$~$>$~$0.77$, \\
$S_{\rm 2-28 ~keV}/S_{\rm 40-700 ~keV}$~$>$~$1.20$, \\
$S_{\rm 2-10 ~keV}/S_{\rm 50-300 ~keV}$~$>$~$1.20$.  

%
%
\begin{figure}
\psfig{figure=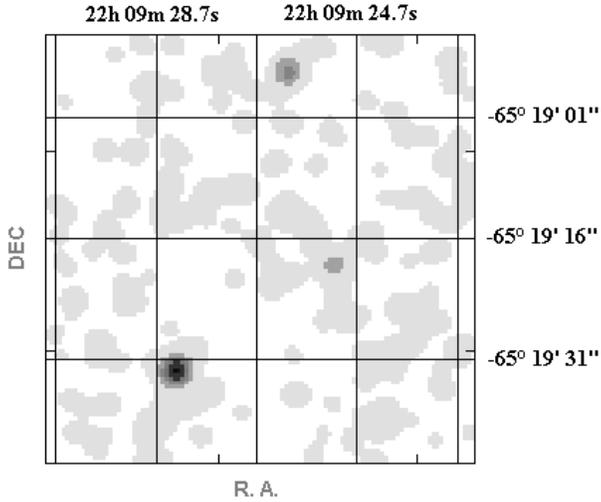,width=7.0cm,angle=-90}
\caption{\chandra\ /ACIS--S image of the field of XRF~020427 
in the  0.3--7 keV energy band accumulated during TOO~1. 
The X--ray afterglow source, 
CXOU J220928.2--651932, is that at the bottom--left side of the image. 
The whole
shown region lies inside the X--ray afterglow error circle determined with the
\sax\ /MECS. 
}
\label{f:acis_image}
\end{figure}

\subsection{Afterglow source detection}
\label{s:image}

The MECS images accumulated over the two \sax\ TOOs in the 1.4--10 keV 
energy range are shown in Fig.~\ref{f:mecs_image}. The X--ray afterglow source 
(1SAX J2209.3--6519) is clearly visible in the
image for the first TOO at an intensity of (5.9$\pm$1.3)$\times$10$^{-3}$ 
cts/s (1.4--10 keV) whereas no significant
signal is detected in the second TOO above a 3$\sigma$ upper limit of 3.0$\times$10$^{-3}$ cts/s.
This implies a
fading by a factor $\ga 2$ from the epoch of the first TOO to that of 
the second one. 

As also reported by Fox (2002)\nocite{Fox02}, the analysis of the \chandra\ images 
reveals
three sources inside the MECS error circle, the brightest of which, 
CXOU J220928.2--651932, shows a decay by a factor $\sim$2.5 from the first to 
the second \chandra\ observation. 
In Fig.~\ref{f:acis_image} we show the 
ACIS--S image accumulated over the first \chandra\ TOO.
 From the sum of their X--ray fluxes
($<$8$\times$10$^{-14}$ $\flux$) we can state that the other two
sources do not significantly contaminate the X--ray flux measurement with the 
MECS 
of the afterglow source. 
The total photon count and flux measured 
during each TOO are reported in Table~\ref{t:tab1} along with those of
the prompt emission.

\subsection{Afterglow spectrum}
\label{s:spectrum}  

\begin{figure}
\vspace{0.2cm}
\psfig{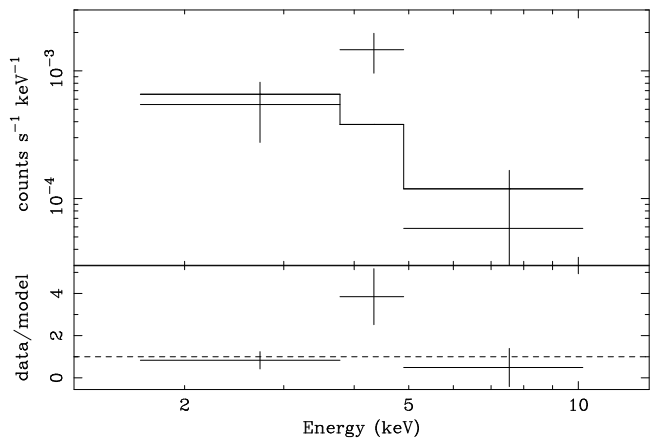}
\vspace{0.2cm}
\caption{MECS TOO~1 spectrum with only three spectral bins in order to
have a sufficient number of counts/bin to
use the $\c2nor$ minimization criterion. The best fit curve with
a power law model absorbed by the average Galactic hydrogen column density along the
line of sight is shown. An excess count above the power law model 
is apparent in the second bin.}
\label{f:aft_spec_mecs}
\end{figure}

\begin{figure}
\vspace{0.2cm}
\psfig{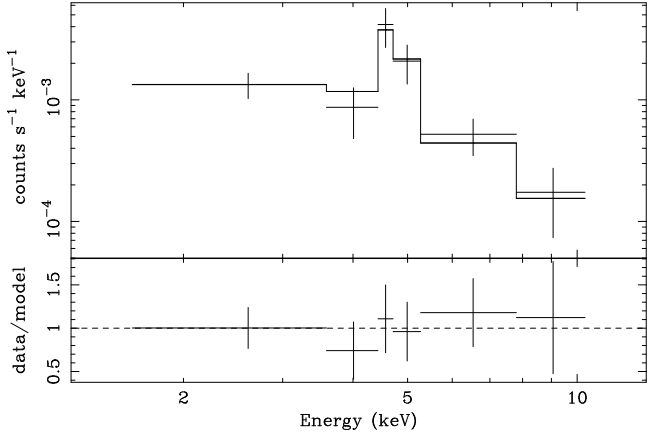}
\vspace{0.2cm}
\caption{MECS TOO~1 spectrum with 6 bins. The shape of the excess in the
2nd bin of Fig.~\ref{f:aft_spec_mecs} is shown. Also shown is the fit
of the spectrum with power law model plus a Gaussian.} 
\label{f:aft_spec_mecs_new_binning}
\end{figure}

\begin{figure}
\vspace{0.2cm}
\psfig{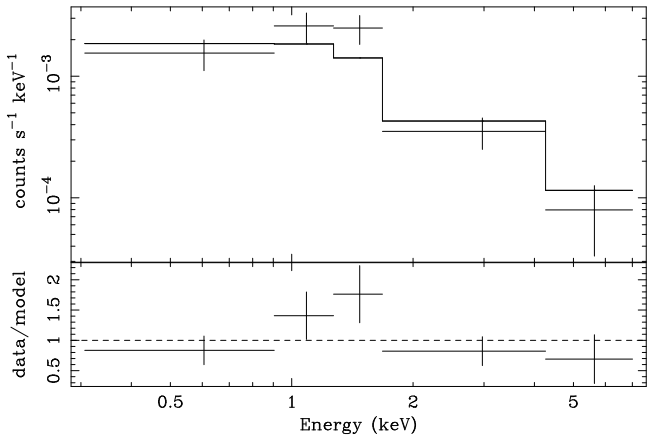}
\vspace{0.2cm}
\caption{Analysis of the ACIS--S TOO~1 spectrum:  
fit with a power law absorbed by
the average Galactic hydrogen column density along the line of sight.} 
\label{f:aft_spec_acis}
\end{figure}

The MECS TOO~1 and ACIS TOO~1 spectra are shown in Figs.~\ref{f:aft_spec_mecs} 
and \ref{f:aft_spec_acis}, respectively. As can be seen from 
Fig.~\ref{f:aft_spec_mecs}, the low statistics do not allow more than 3 bins 
with a sufficient number of counts/bin to allow the use of 
the $\c2nor$ statistics. 
The fit of this spectrum with an absorbed ($N_{\rm H}^G = 2.9 \times 10^{20}$~cm$^{-2}$) 
power law (see Table~\ref{t:tab2})
is only marginally
acceptable ($\c2nor$/dof = 5.1/1, null hypothesis probability, NHP, = 0.02), 
which could be due to either an excess count in the second energy bin 
(3.8--5 keV) or to a shortage count in the first bin (1.6--3.8 keV).
Assuming the latter possibility, we performed a number of
fits with the column density fixed at different values and the 
power law parameters free to vary. The result was that the best 
description of the data ($\c2nor$/dof = 1.1/1) is obtained with N$_H = 2.8\times
10^{23}$~cm$^{-2}$ and photon index $\Gamma = 7.3^{+2.9}_{-2.1}$.  We note that such an 
high value of $\Gamma$ has never been observed in GRBs or XRFs 
(e.g. \cite{Frontera03})
and it would be very difficult to find a physical interpretation for it. 
Moreover, the high value of N$_H$, which
could be due, e.g., to a very dense environment surrounding the source, 
is largely inconsistent with that inferred from the WFC spectral measurements of 
the prompt emission (see Sect. 3.1). 
If, alternatively, we try to fit the excess in the second bin with a Gaussian,
the centroid of this is at
$4.7_{-0.7}^{+1.7}$~keV. (However, this can be interpreted with extreme caution). \\
We evaluated the chance probability of this result by means of numerical
simulations. We simulated 1000 spectra by folding an 
absorbed power law photon spectrum ($\Gamma=2.0$, $N_{\rm H} = N_{\rm H}^G$) 
with the MECS response function and adding Poisson noise. We also took
into account the MECS background. Each simulated spectrum was rebinned
exactly like the measured spectrum (3 bins). We found that in 11 cases out of 
1000 the fit with an absorbed power law gave a $\c2nor/{\rm dof} \ge 5.1/1$,
corresponding to a chance probability of 0.011 that the observed excess in the 
second bin is due to chance.\\
By increasing the number of spectral bins the MECS TOO~1 spectrum
has the shape shown in Fig.~\ref{f:aft_spec_mecs_new_binning}, which
actually shows that the excess is  concentrated around 4.5--5 keV. By
subdividing the MECS TOO~1 observation in two parts, it turns out that the
feature is observed only in the first part. Also, it is found in both spectra
of the MECS telescope units, thus excluding that it is due to the instability 
of one of the detectors. 

Fig.~\ref{f:aft_spec_acis} shows the ACIS TOO~1 spectrum. No special
feature is apparent, even if
the fit with a power law absorbed by the average Galactic 
column density $N_{\rm H}^G$ (see Table~\ref{t:tab2}) is not completely satisfactory ($\chi^2/{\rm dof}
= 4.7/2$, NHP=0.09). Leaving $N_{\rm H}$ free, we obtain a better fit
($\chi^2/{\rm dof} = 0.6/2$), with a best fit value of $N_{\rm H}$ 
higher than $N_{\rm H}^G$ but poorly constrained. 
The photon index value is consistent with that derived from the spectral analysis 
of
the first \sax\ / MECS observation. 
Also, the best fit parameters of the power law model for the prompt emission
spectrum (Sect.~3.1 and Table~2) are consistent, within statistical 
uncertainties,
with those derived for the afterglow emission.

%
%
\begin{center}
\begin{table}[t!]
\caption{Log of the spectral fits of the last pulse of the prompt emission and 
of the 
X--ray afterglow with a photoelectrically absorbed power law.
 The energy band and exposures are the same reported in Table~\ref{t:tab1}. 
The quoted uncertainties are at 90\% confidence level.}
\begin{tabular}{lcccc}
\hline
\hline Observation & N$_H$ & $\Gamma$ & $\cqr$ \\
      & (10$^{21}$ $\rcm2$)  &   &  \\
\hline
prompt emission & [0.29]  & 2.22$_{-0.25}^{+0.31}$ & 7.4/8 \\
(second pulse) \\
\hline
MECS TOO~1 & [0.29] & 2.0$_{-1.1}^{+2.2}$ &  5.1/1\\
\hline
ACIS--S TOO~1 & [0.29]  & 1.5$_{-0.5}^{+0.5}$   & 4.7/2 \\
\hline
\hline
\end{tabular}
\label{t:tab2}
\end{table}
\end{center}

\subsection{Afterglow time behavior}
\label{s:fading}

\begin{figure*}
\centerline{\psfig{figure=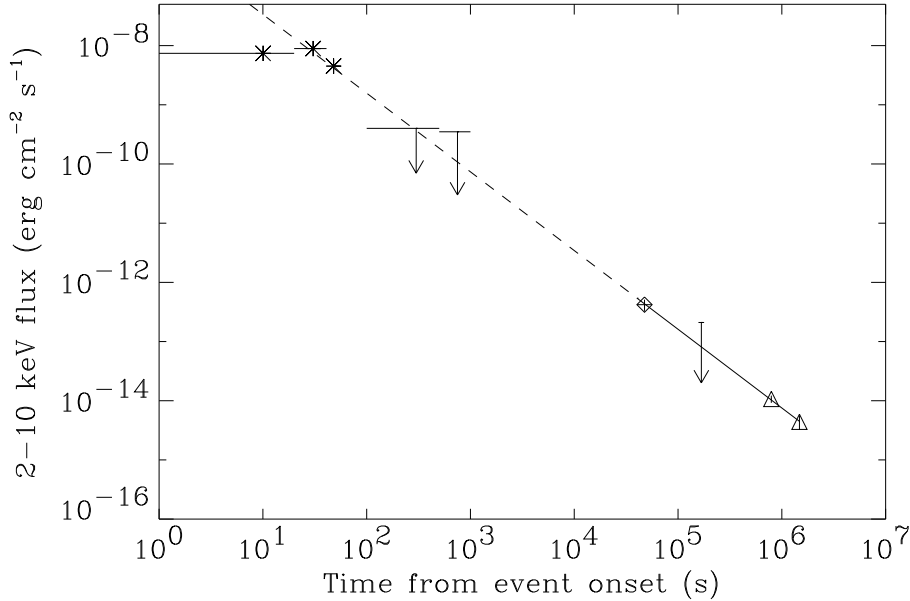,width=12.5cm}}
\caption{Temporal evolution of the 2-10 keV flux of XRF~020427 from the
prompt to afterglow emission. The first three points ({\it asterisks}) 
correspond to the average flux measured during the prompt emission, the 
third point ({\it diamond}) to the average flux measured by the MECS 
during the first \sax\ TOO, and the last two data points ({\it triangles}) 
to the average fluxes measured by ACIS--S during the two \chandra\
TOOs. The  two  upper limits around  500 s were obtained with the \sax\ WFC, 
while the other with MECS. The upper limits are at 3$\sigma$ confidence level. 
The continuous line is the best fit power law to the MECS + ACIS--S data 
(see text), the dashed line is its extrapolation back to the time of the
prompt emission.}
\label{f:aft_lc}
\end{figure*}

The 2--10 keV light curve of the X--ray afterglow is shown in 
Fig.~\ref{f:aft_lc}. The figure also shows the time profile of the prompt 
emission in three time bins and the WFC flux upper limits to the
XRF emission from 100 to 1000 s after the event onset, 
derived assuming the spectral model which best fits the
average spectrum of the prompt emission.

The three points corresponding to MECS TOO~1 and ACIS--S data are well fit
by a power law ($F(t) \propto t^{-\delta}$) with a decay index 
$\delta$=1.30$_{-0.09}^{+0.10}$. In addition, the back extrapolation of the 
afterglow decay law to the time of the prompt emission  is
consistent with the last two WFC points, corresponding to the second pulse
of the prompt emission, and with the WFC upper limits. Even though we cannot exclude
other possibilities, this extrapolation is consistent with the indication, found for several 
GRBs (e.g. Frontera et al. 2000b), 
that the
afterglow emission starts during the late part 
of the prompt emission.
By including in the fit the two WFC points, we obtain 
$\delta$=1.33$_{-0.03}^{+0.02}$. 

%
%

%
%

%
%

\section{Discussion}
\label{s:disc}

The nature of XRFs is still an open issue. As discussed in Sect.~1, even if 
the evidence is mounting that XRFs are a 
sub-class of GRBs, 
it is still debated whether observational properties (mainly X--ray richness 
and peak energies of the order of a few keV or less) are due to 
intrinsic properties of the fireball 
(e.g. very low fireball bulk Lorentz factors due to high baryon loading), to a 
very large distance (z$>$5), or to collimated GRBs seen at large off axis angles.

XRF~020427 offers us the opportunity to fill in some details about
the XRF class properties 
and constrain models of X--ray flashes.
Indeed, this is the first  XRF for which a joint study of the prompt and 
X--ray afterglow emission, 
up to about 
17 days after the main event, is presented. We discuss below the main 
distinguishing features of XRF~020427 and their theoretical implications.

\subsection{Common and distinguishing features of XRF~020427 with respect 
to GRBs}

The lower limit (1.2)
to the ratio between the 2--28 keV and 40--700 keV fluences is significantly 
higher than the average value found for normal GRBs 
(e.g. \cite{Feroci01,Barraud03}), but it is lower than the values 
($\sim$2.8--3.5) found for the \sax\ most X--ray rich  GRBs, namely GRB~981226
(\cite{Frontera00a}), GRB~990704 (\cite{Feroci01}) and GRB~000615 
(\cite{Maiorano03}).
Still, XRF~020427 shows a much softer prompt 
emission spectrum than these events, with $E_{\rm p} < 5.5$~keV 
(see 
Sect.~3.1). Also an association with the
simultaneous ionospheric disturbance (see Sect.~\ref{s:obs}) 
favors a very low $E_{\rm p}$. The $E_{\rm p}$ upper limit is consistent with the 
low--energy tail of the $E_{\rm p}$ distribution found for a sample including 
both 
GRBs and XRFs (\cite{Kippen03}), but, as it happens for very 
few XRFs (\cite{Kippen03}), it is inconsistent 
with the $E_{\rm p}$ distribution of GRBs as a function of their duration.  

XRF~020427 shows X--ray afterglow emission, like most ($\sim$90\%) of the 
accurately localized and followed--up GRBs, including the most X--ray rich 
(\cite{Frontera00a,Feroci01,Maiorano03}). Remarkably, the spectral and 
temporal properties of the XRF afterglow are similar to those of GRBs. 
The power law 
photon index, even though poorly determined ($\Gamma = 2.0_{-1.1}^{+2.2}$), 
is not in contradiction with the mean value found for GRBs (1.93$\pm$0.35, 
\cite{Frontera03}). 
The power law decay index $\delta$ = 1.30$_{-0.09}^{+0.10}$ is
nearly coincident with the average value found for GRBs 
(1.33$\pm$0.33, \cite{Frontera03}), and the
back extrapolation of the decay law to the time of the primary event
is consistent with the late part of the prompt emission, as found for
many GRBs (e.g. \cite{Frontera00b}).

Both the (extrapolated) 2--10 keV fluence of the afterglow in the time interval 
from 60~s to 10$^{6}$s
($5.4^{+1.2}_{-1.2} \times 10^{-7}$~$\flue$) and its ratio ($1.46 \pm 0.32$) to 
that of the prompt emission in the same energy band are in the range of 
values found for GRBs (\cite{Frontera00b}).
Nevertheless, the lower limit to the ratio (1.13) between the 2--10 keV afterglow 
fluence and the gamma--ray (40--700 keV) fluence of the prompt emission is 
much greater than the values
typically found for GRBs, which range from few percent to $\sim$30\% 
(e.g. \cite{Frontera00b}). 
This could indicate that XRFs are events in which fireball internal shocks, 
thought to be responsible for most of the GRB prompt emission, are poorly 
efficient, whereas the external shock(s) generating the afterglow emission 
work with the same efficiency as for normal GRBs.

%
%

\subsection{Redshift and energetics of XRF~020427}
\label{s:redshift}

As discussed in Sect.~2, a redshift upper limit 
of 2.3 has been estimated for the candidate host galaxy of XRF~020427 
(\cite{Dokkum03}). From the X--ray data we derive further 
redshift constraints for this XRF event. 

A possible estimate derives from the relationship for GRBs
between the {\it intrinsic} peak energy 
of the $E F(E)$ spectrum, 
$E_{\rm p}^{\rm i}$
(not to be confused with the {\it
observed} peak energy $E_{\rm p}$),
and the isotropically radiated energy, 
E$_{\rm rad}$, (\cite{Amati02}). This relationship may very well
apply to XRFs, as exemplified by XRF~020903 
and XRF~030723, if the estimates of their redshift are true
(\cite{Lamb03}).
From the derived upper limit to the {\it observed} peak energy,
$E_{\rm p} < 5.5$ keV (see 
Sect.~\ref{s:prompt}), the $E_{\rm p}^i$ vs. $E_{rad}$ relation is 
satisfied for 
z$\la$0.1--0.2, with $E_{\rm rad} < (0.5-1.5) \times 10^{50}$~erg.

Also from the peak luminosity versus variability correlation found 
by Vetere et al. 
(2003)
\nocite{Vetere03} we can derive a redshift estimate. Using a sample 
of GRBs simultaneously detected by the \sax\ WFC and GRBM with known redshifts, 
Vetere et al. (2003) found that the gamma--ray (40--700 keV) peak luminosity, 
L$_{\gamma}$, is related to the X--ray (2--28 keV) light curve 
variability, 
V$_X$,  according to L$_{\gamma}\propto V_X^{2.48\pm 0.54}$, similar 
to the correlation
found by Reichart et al. (2001)\nocite{Reichart01} for the peak 
luminosity -- gamma--ray variability 
V$_{\gamma}$ (L$_{\gamma}\propto V_\gamma^{3.3\pm 1.0}$). For XRF~020427, we
evaluated V$_X$ following Vetere et al. (2003), and we derived the
gamma--ray (40--700 keV) peak flux from the measured 2--28 keV peak
flux assuming the average spectrum derived in Sect.~\ref{s:prompt}. 
The redshift 
values which satisfy the L$_{\gamma}$ vs. V$_X$  relation are in the range 
0.3--0.9. 

Assuming the highest redshift upper limit ($z = 2.3$), we get 
the following upper limits to $E_{\rm p}^i$ and $E_{rad}$: $E_{\rm p}^i <16.5$~keV 
and 
$E_{rad}< 5.7 \times 10^{52}$~erg, where $E_{rad}$ was evaluated following
the method described by Amati et al. (2002), assuming as spectral
shape a Band law (\cite{Band93})
with $\alpha= -1$, $E_0 = 5.5$~keV and $\beta = -2.10$ 
(see Sect.~\ref{s:prompt}) 
and a flat Friedman-Robertson--Walker cosmological model with $H_0 = 
65$~km~s$^{-1}$~Mpc$^{-1}$, $\Omega_m = 0.3$, $\Omega_\Lambda = 0.7$
(e.g. \cite{Carroll92}).

\subsection{Testing the off axis jet scenario}

From the redshift upper limits derived above it follows that the 
extreme 
softness of XRF~020427 is not a redshift effect.
A possible explanation is
that XRF~020427 involves a collimated jet seen at large off axis angle,
as proposed
by various authors (e.g. \cite{Granot02,Rossi02,Yamazaki02}). 

Assuming a uniform jet with half opening
angle $\Delta\theta$ and Lorentz factor $\gamma$, it can be shown that the 
value of $E_{\rm p}$ measured at a viewing angle $\theta_v$ with respect 
to the jet axis is constant if 
$\theta_v < \Delta\theta$ but decreases by a factor 
$\delta \sim 2\gamma/[1 + \gamma^2(\theta_v - \Delta\theta)^2]$ 
for $\theta_v > \Delta\theta$ (e.g. \cite{Yamazaki02}). 
In structured jet scenarios (e.g. \cite{Rossi02}), $E_{\rm p}$ is a function of 
$\theta_v$ also for $\theta_v < \Delta\theta$ in a way that 
depends on the beam profile assumed. Assuming a uniform jet emission,
from the derived upper limit of the observed peak energy $E_{\rm p}$ and the above formula 
we get $\theta_v - \Delta\theta \ga 30\deg$, if  XRF~020427 
is a normal GRB with an on axis $E_{\rm p}$ value of 200 keV, 
a typical
Lorentz factor $\gamma$ of 150 (\cite{Frontera00b}), and a redshift $z<0.9$
inferred above. 
 
However the off axis jet models foresee that, for large off axis observers 
(i.e., for
$\theta_v > \Delta\theta$ ) the afterglow light curve should be 
characterized
by a sharp (case of a uniform jet) or smooth (case of a 
structured jet) 
rise, a peak and a subsequent
power law decay (e.g. \cite{Granot02,Rossi02}). 
This behavior is not found in our data: the afterglow flux monotonically 
decays as a power law from the second half of 
the prompt emission, 25 s from the event onset, up to the end of the
second \chandra\ TOO, 17 days after the event onset. The only possibility to
meet the expectations of these models is that the afterglow peak is achieved 
during the second pulse of the prompt emission.  
However, for viewing angles greater than the lower limit estimated above, the peak 
is expected to occur at much later times (e.g. \cite{Granot02,Rossi02,Dalal02}). 
Thus, the interpretation of the X--ray data in terms of an off axis observation
of XRF~020427 is difficult.

If we abandon the off axis hypothesis, we can set a lower limit to
the afterglow light curve break time $t_b > 17$~days by assuming that the decay
is indeed monotonic from the prompt emission up to the end of the
second \chandra\ TOO. By using the relation 
$t_b = 6.2(1+z)(E_{52}/n)^{1/3}(\Delta\theta/0.1)^{8/3}$ hr (\cite{Sari99a}) and
assuming {\it n} = 1 cm$^{-3}$ and z$<$0.9 (as discussed above), 
which implies
$E_{52}$$<$0.5,
we derive a lower limit to the jet opening angle of $\Delta\theta > 23.5\deg$.
We note that 
higher
values of $n$ would increase the lower limit to $\Delta\theta$ .
The value of $n$ is expected
to be about 1--10 cm$^{-3}$ for events occurring within a galaxy outside 
star forming regions and, in principle, could be as high as
 10$^5$--10$^6$ cm$^{-3}$ for events occurring in
dense star forming regions (e.g. \cite{Bottcher99,Ghisellini99}). 
However, the estimated values of $n$ for several GRBs never exceed
$\sim$50 cm$^{-3}$ (e.g. \cite{Panaitescu01}). In addition, the lack of
evidence of a N$_H$ higher than the average galactic one along the
line of sight, the absence of a break in the afterglow light curve,
which is expected in case of an early transition from relativistic to 
sub--relativistic expansion due to a very dense circum--burst material
(e.g. \cite{Huang98,Dai99,Zand01}), and
the blue color of the host galaxy favor the hypothesis of a low or moderate 
density of the medium 
surrounding XRF~020427.
Thus, $n=1$ is a reasonable assumption.

\subsection{Testing the fireball baryon loading}

Given that there is no evidence supporting the observation of a narrow 
relativistic jet seen from large viewing angles with respect to the jet axis,
other scenarios need to be investigated.

In internal shock models, it is found (e.g. \cite{Zhang02,Mochkovitch03}) 
that a low value 
of $E_{\rm p}$ is the consequence of a clean fireball with a high $\gamma$, because 
in this case the shocks are expected to be less efficient due to the lower 
contrast of the Lorentz factor between two colliding shells and to the
greater distances from the central engine, and thus lower magnetic field, at 
which the shocks occur. 

In the external shock scenario, a natural way to obtain a low value 
of E$_{\rm p}$ is 
a dirty fireball, i.e. a fireball with a higher baryon loading and thus a 
lower Lorentz factor $\gamma$ with respect to normal 
GRBs (e.g. \cite{Dermer99}). In this scenario $E_{\rm p}$ is positively 
correlated with $\gamma$. 

Assuming that the second pulse of the 
XRF is the peak of the afterglow flux,
one can derive the Lorentz factor from the afterglow rise time 
(with respect to the 
GRB onset), $t_0$, 
the number density, $n$, of the ambient medium and the afterglow energy, 
$E_{aft}$, released in the shock, according to
 Sari \& Piran (1999)\nocite{Sari99b},
$\gamma = 240(E_{aft}/10^{52}erg)^{1/8} n^{-1/8} (t_0/10s)^{-3/8}$. 
Taking into account that $E_{aft}$ ranges  
from $1.2\times10^{50}$  to $1.1\times10^{52}$~erg, depending on $z$ in the
range from 0.1 to  0.9 (see Sect.~4.2) and $t_0 = 25$~s (corresponding
to the second pulse), we get 
$\gamma > 195$. 
This value is of the same order of 
magnitude as normal GRBs (e.g. Frontera et al. 2000b\nocite{Frontera00b})
and favors the hypothesis that the prompt emission 
is generated by internal shocks in a 
clean fireball. In this computation we assumed $n$ = 1, which is 
 a reasonable value for this source as discussed in Sect. 4.3 .
We note that assuming a value of $n$ as high as 100~$cm^{-3}$ 
would reduce by a factor of $\sim$2
the estimated
lower limit to $\gamma$.

\section{Conclusions}

Up to very recently, the study of XRFs and their comparison with GRBs was based on the
prompt emission only (e.g. Heise et al. 2003, Kippen et al. 2003).
In this paper we presented for the first time a detailed and joint analysis of 
the 
prompt and afterglow X--ray emission
of an X--Ray Flash (XRF~020427). The prompt event emission was observed
with the \sax\ WFC and GRBM, while the afterglow emission was measured with
the \sax\  MECS and \chandra\ ACIS instruments. 

The prompt emission spectral analysis shows that XRF~020427 belongs to the class of
very soft events ($E_{\rm p}<5.5$~keV). The spectral hardness and the duration 
of the event are inconsistent with the correlation found between these
two quantities for normal and X--ray rich GRBs. 

The X--ray afterglow intensity, spectrum and temporal decay are similar 
to those of normal GRBs. Also, the extrapolation of the afterglow decay
law 
back to the time of the primary event is consistent with the flux of the
late prompt 
emission, as commonly found for normal GRBs (Frontera et al. 2000). 
A marginally significant excess on the continuum
 was found 
at $\sim 4.5-5$~keV in the MECS spectrum of the afterglow source during the 
first \sax\ observation.

The XRF properties do not seem a consequence of a high GRB redshift.
By assuming that the relationships between
intrinsic 
peak energy and total radiated energy (\cite{Amati02}) 
and between peak luminosity and
X--ray variability (\cite{Vetere03})
found for normal GRBs hold also for XRFs, we find $z \la$0.9. 
These estimates further constrain the upper limit of 
$z<2.3$ inferred for the host 
galaxy redshift.

We have investigated possible scenarios to interpret the derived XRF emission
properties.
Assuming a homogeneous off axis jet model (e.g. \cite{Granot02}), the measured
peak energy of the $EF(E)$ spectrum of the prompt emission could be interpreted
as due to a large off axis 
viewing angle of the jet (at least  $\sim23\deg$ larger than the jet 
opening angle ).
However the afterglow light curve shows a monotonic decay from the
prompt emission up to 17 days after the event onset, in contrast to
the predictions of this model  (e.g. \cite{Granot02}).

Also a highly baryon loaded fireball scenario appears problematic 
given the high lower limit for the Lorentz factor, 
in contrast to the predictions of this model ($\gamma << 100$, 
e.g., \cite{Dermer99,Huang02}) even by assuming a medium density 
substantially
higher than typically observed for GRBs.
Instead, the very low value of the peak energy and the high lower limit for
$\gamma$ point to a clean fireball with high Lorentz factor, in which the
XRF is produced by internal shocks with a very low efficiency due to the
small contrast of $\gamma$ between colliding shells (e.g. \cite{Mochkovitch03}).

\acknowledgements 
This research was partly supported by the
Italian Space Agency (ASI). We wish to thank the teams of the \sax\
Operative Control Center and Scientific Data Center for their
efficient and enthusiastic support to the GRB alert program. We also 
thank the \chandra\ Data Archive Operations Group for their very useful work.
Finally, we thank 
Ryo Yamazaki for useful discussion and Franco Giovannelli for the 
thorough 
critical reading of the paper and useful comments and
suggestions.

\end{document}